\documentclass[conference,compsoc]{IEEEtran}
\IEEEoverridecommandlockouts

\newcommand{\typeA}[1]{\cellcolor{white}{#1}}
\newcommand{\typeB}[1]{\cellcolor{ForestGreen!20}{#1}}
\newcommand{\typeC}[1]{\cellcolor{Tan!20}{#1}}
\newcommand{\typeD}[1]{\cellcolor{Gray!35}{#1}}

\newcommand{\sqTypeB}[1][ForestGreen!30]{\textcolor{#1}{\ensuremath\blacksquare}}
\newcommand{\sqTypeC}[1][Tan!30]{\textcolor{#1}{\ensuremath\blacksquare}}

\usepackage{url}
\usepackage{cite}
\usepackage{amsmath,amssymb,amsfonts}
\usepackage{algorithmic}
\usepackage{graphicx}
\usepackage{textcomp}
\usepackage{xurl}
\usepackage[colorlinks=true,urlcolor=black]{hyperref}
\usepackage[table,dvipsnames]{xcolor}
\usepackage{comment}
\usepackage{cleveref}
\usepackage{multirow}
\usepackage{rotating}
\usepackage{enumitem}
\usepackage{wrapfig}
\usepackage{bmpsize}
\usepackage[]{footmisc}
\def\BibTeX{{\rm B\kern-.05em{\sc i\kern-.025em b}\kern-.08em
    T\kern-.1667em\lower.7ex\hbox{E}\kern-.125emX}}

\begin{document}

\title{SoK: A Systems Perspective on Compound AI Threats and Countermeasures}


\author{\IEEEauthorblockN{ 
Sarbartha Banerjee\IEEEauthorrefmark{1}\IEEEauthorrefmark{2}, \;
Prateek Sahu\IEEEauthorrefmark{1}\IEEEauthorrefmark{2}, \;
Mulong Luo\IEEEauthorrefmark{2}, \\
Anjo Vahldiek-Oberwagner\IEEEauthorrefmark{3}, \;
Neeraja J. Yadwadkar\IEEEauthorrefmark{2}, \;
and  Mohit Tiwari\IEEEauthorrefmark{2}\IEEEauthorrefmark{4}} \vspace{0.3em}
\IEEEauthorblockA{\IEEEauthorrefmark{2}The University of Texas at Austin \;\; \IEEEauthorrefmark{3}Intel Labs \;\; \IEEEauthorrefmark{4}Symmetry Systems}
}



\maketitle

\begingroup\begin{NoHyper}\renewcommand\thefootnote{\IEEEauthorrefmark{1}}
\footnotetext{Sarbartha Banerjee and Prateek Sahu are equal contributors.}\end{NoHyper}

\begin{abstract}


Large language models (LLMs) used across enterprises often use proprietary models and operate on sensitive inputs and data. 
The wide range of attack vectors identified in prior research—targeting various software and hardware components used in training and inference—makes it extremely challenging to enforce confidentiality and integrity policies.

As we advance towards constructing compound AI inference pipelines that integrate multiple large language models (LLMs), the attack surfaces expand significantly. Attackers now focus on the AI algorithms as well as the software and hardware components associated with these systems. While current research often examines these elements in isolation, we find that combining cross-layer attack observations can enable powerful end-to-end attacks with minimal assumptions about the threat model.
Given, the sheer number of existing attacks at each layer, we need a holistic and systemized understanding of different attack vectors at each layer.

This SoK discusses different software and hardware attacks applicable to compound AI systems and demonstrates how combining multiple attack mechanisms can reduce the threat model assumptions required for an isolated attack. 
Next, we systematize the ML attacks in lines with the $Mitre\ Att\&ck$ framework to better position each attack based on the threat model.
Finally, we outline the existing countermeasures for both software and hardware layers and discuss the necessity of a comprehensive defense strategy to enable the secure and high-performance deployment of compound AI systems.

\end{abstract}


\section{Introduction}

AI-powered applications like chatbots~\cite{ChatGPT,Gemini,Copilot}, and ML tools like autonomous driving~\cite{Autopilot} and OCR~\cite{OCRSoftware}, have become widespread due to advances in neural networks and transformers. Recent developments in large language models (LLMs) with billions of parameters support tasks like program and image generation. These production-ready models train on massive data, often containing sensitive information, and utilize proprietary architectures running on expensive compute resources, making them targets for adversaries.
Literature in ML security and privacy have explored several attacks~\cite{shokri2017membership,tramer2016stealing,milli2019model,tian2022comprehensive,band_util_2} and developed efficient defenses~\cite{banerjee2020sesame,banerjee2023triton,juuti2019prada,bourtoule2021machine,borgnia2021strong}, but many focus individually on algorithms or platforms, ignoring the complex interactions in modern AI systems.
A recent study~\cite{carlini} also highlighted that privacy-preserving models can leak data when deployed alongside other software components, challenging existing security guarantees.

\begin{figure}
    \centering
    \includegraphics[trim={0.9cm 0.4cm 1cm 0.5cm},clip,width=1\linewidth]{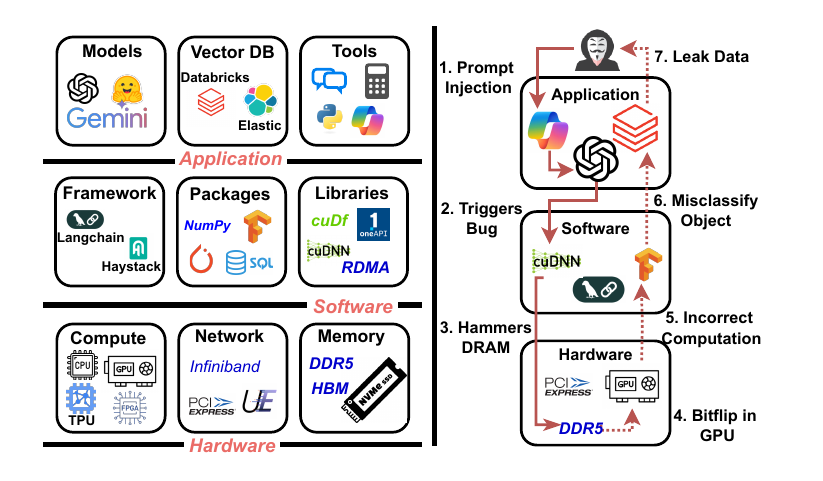}
    \vspace{-1.5em}
    \caption{\bf Application, software and hardware layers for compound AI.
    Example shows cross-layer components can be exploited to leak data.
    }
    \label{fig:AI-Stack}
\end{figure}

\begin{figure*}
    \centering
    \includegraphics[trim={0 0 0 0},clip,width=1\linewidth]{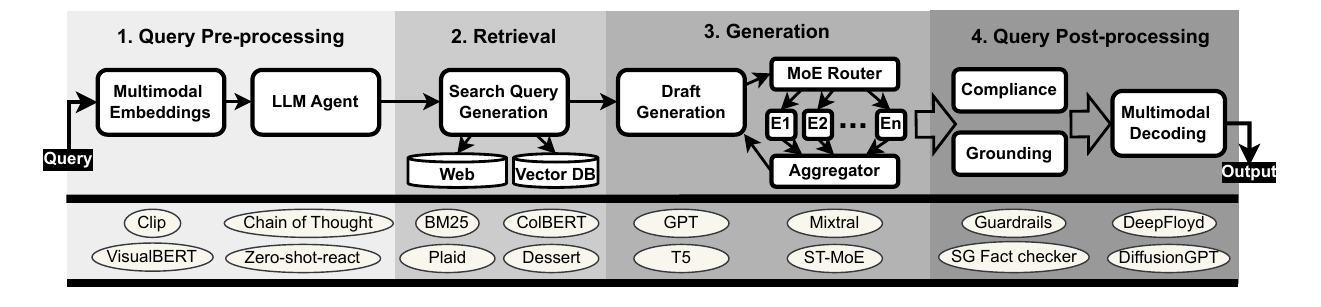}
    \vspace{-1.5em}
    \caption{\bf A Compound AI Pipeline begins with Query pre-processing to refine the input query and feed to an LLM agent. 
    The agent extracts knowledge in the Retrieval stage, generates a draft response, and fills information generated from MoE experts in the Generation step.
    This response goes through compliance and fact checks Query post-processing step to finally generate a multi-modal output.
    }
    \label{fig:AI-Pipeline}
\end{figure*}

Emerging compound AI systems~\cite{compound-ai-blog} are created by integrating multiple AI models with numerous software components, all deployed across distributed hardware.
~\Cref{fig:AI-Stack}-left provides a cross-stack view of a compound AI system:
The \textit{application} layer comprises of multiple AI models, vectorized databases for knowledge storage, and associated tools used by AI agent models.
The \textit{software} layer includes frameworks like Langchain~\cite{Langchain} to design AI pipelines, packages like Pytorch~\cite{Pytorch} to design models, and libraries like cuDNN~\cite{Cudnn} to interface with devices.
Finally, the \textit{hardware} layer comprises of compute units like CPUs, GPUs, and TPUs along with memory and network components.
While literature have proposed attacks across different AI/ML models and systems, vulnerabilities and attack surfaces across different components in these layers can be composed to create new attack vectors or relax the threat model assumptions for existing attacks.

The complexity of layered software, distrustful entities, and diverse hardware creates a broad attack surface, highlighting the need for a systematization of existing literature. The volume of proposed attacks and defenses across different stack layers makes it challenging to grasp threat models and associated defenses. Heterogeneous hardware -- CPUs, GPUs, ASICs, FPGAs -- introduces further risks, including digital and physical side-channel attacks, that could expose sensitive data or model parameters. Device ownership between cloud and edge deployments, adds complexity to the trust landscape, expanding the overall attack surface.
~\Cref{fig:AI-Stack}-right enumerates an example of how complex cross-layer observations can lead to an end-to-end attack.
An attacker performs a prompt injection attack that can exploit a software library bug, eventually triggering repeated memory access to adjacent memory rows containing model parameters. 
Repeated access can result in Rowhammer~\cite{mutlu2019rowhammer}, that changes the weights tensors due to bit-flips. 
These tampered weights can change the model prediction and misclassify objects which eventually generates incorrect response. 

However, no prior work has systematized the impact of vulnerabilities in system software (e.g., frameworks, packages, libraries) or hardware (e.g., side-channels, fault injection) in a heterogeneous deployment platforms. We categorize these as \textit{system attacks} and focus on organizing them with a view to the complex threat models emerging from compound AI systems.
Algorithmic attacks~\cite{tian2022comprehensive,tramer2016stealing,shokri2017membership,memorization}, that target ML algorithms and training data in the application layers, are well-studied by prior SoKs~\cite{sok1,sok2,sok3}. Our work also explores attacks at the software and the hardware layers that attackers can use for composing cross-layer widgets to perform an end-to-end attack. 
Understanding the threat model assumptions and attacker capabilities can inform system designers of cross-stack attack composition as well as define defense mechanisms at the software and hardware layer. This also informs algorithm designers of system and hardware impacts, and help build robust models.

In this paper, we focus on different types of \textit{system} attacks and defenses in a compound AI system. Specifically, we make the following contributions --
\begin{enumerate}[leftmargin=*]
    \item We explore a range of software, and hardware vulnerabilities, as well as the defenses impacting AI models and study how they affect compound AI pipelines. To the best of our knowledge, this is the first effort to methodically categorize system attacks and defenses for AI systems.
    \item We focus on the cross-stack vulnerabilities introduced by the emergence of compound AI pipelines in this domain and systematize existing attacks and vulnerabilities in a well-known cyber-security framework -- $Mitre\ Att\&ck$~\cite{mitre}. 
    This can be used a foundation for threat model developments and methodologies for individual defenses. 
    \item We explore case-studies which utilize cross-stack vulnerabilities to mount an end-to-end attack. We discuss the learning and gaps of current security practices and bring out open research questions that will drive us to build better and safer AI applications.  
\end{enumerate}

The rest of the paper is organized as follows. \Cref{sec:background} introduces emerging compound AI pipelines and related AI/ML security literature. \Cref{sec:threat_model} discusses the assets, trust entities and threat models under study. \Cref{sec:sw_att_def} and \Cref{sec:hw_att_def} explores various software and hardware attacks and defenses of existing AI/ML systems. \Cref{sec:systematize} works on systematizing the attacks in a well formed framework and explores new attack paths that are possible within the paradigm of modern compound AI pipelines. Finally, \Cref{sec:discussion} addresses key learning and gaps of current security practices
before concluding in \cref{sec:conc}.
\section{Background}
\label{sec:background}

Language models are popular for next token prediction, given an input prompt. With the prolific use of accelerated hardware (GPUs, TPUs~\cite{tpu} etc.), such models have grown to billions of parameters 
with the capability to understand complex structures and provide high accuracy responses -- making large language models (LLMs).

\noindent {\bf Compound AI:} The growing demands of LLMs and their widespread use bring new system and design challenges. Modern AI inference pipelines often involve multiple ML and LLM models working together, from prompt engineering to token generation and AI-safety checks. The complexity increases with the use of multi-modal inputs (e.g., Gemini\cite{Gemini}) and fine-tuned mixtures of experts (e.g., GitHub Copilot\cite{gitCopilot}, Amazon Q Developer\cite{amazon_q}, MistralAI\cite{mixtral}), resulting in Compound AI systems\cite{compound-ai-blog}.

\begin{figure*}
    \centering
    \includegraphics[trim={0.7cm 0.2cm 0.7cm 0.2cm},clip,width=1\linewidth]{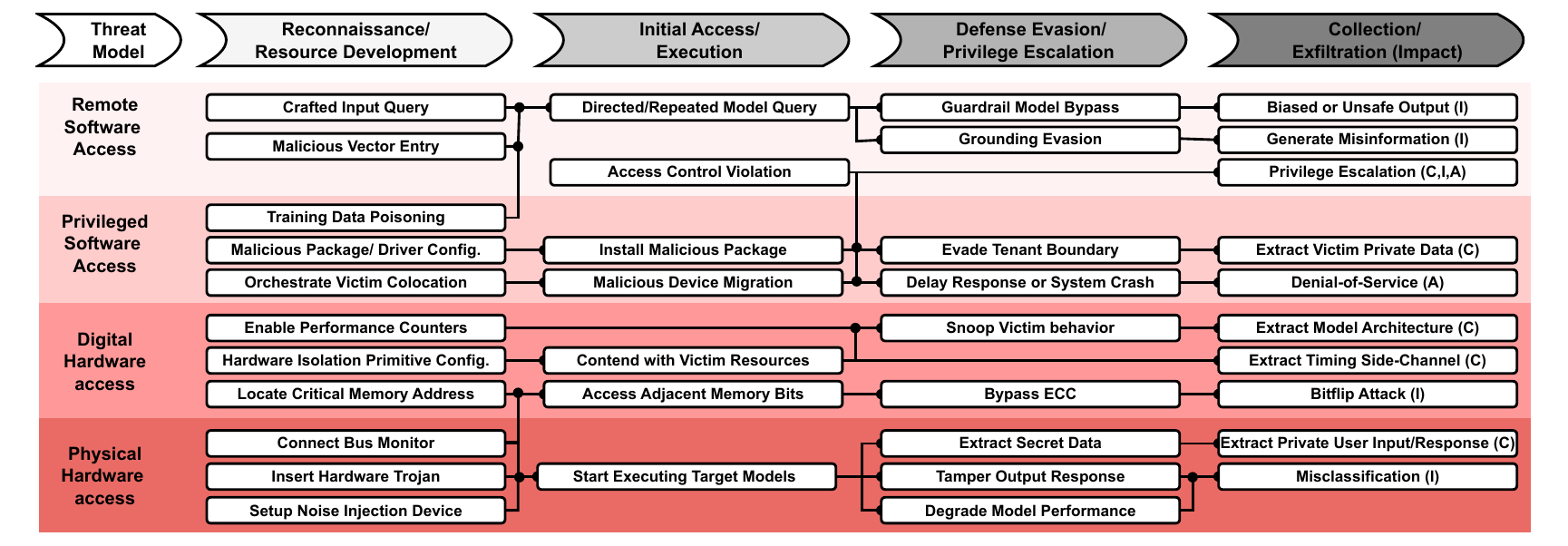}
    \vspace{-1em}
    \caption{\bf Positioning each attack widget in the Mitre Attack framework. 
    The different colors specify the attacker capability ranging from remote access (weakest adversary) to physical access (strongest adversary). 
    The attack steps through different steps. The impact denotes the security policy violation [C = Confidentiality, I = Integrity, A = Availability]
    }
    \label{fig:mitre_tm}
\end{figure*}

These pipelines integrate diverse software and ML/LLM components to optimize resource usage and improve accuracy. In a typical workflow shown in \cref{fig:AI-Pipeline}, a user query, potentially containing text, images, and videos, is processed to generate embedded context~\cite{clip}. Optimizations like few-shot training or chain-of-thought reasoning are applied, and facts may be retrieved using a Retrieval Augmented Generator (RAG)~\cite{Copilot}. The query is then sent to an LLM for token generation, often using a mixture-of-experts approach. Once generated, responses undergo grounding and safety checks before being presented back to the user.
This complexity requires sophisticated software orchestration to manage model batching, data transfer, and resource allocation across heterogeneous platforms, going beyond traditional GPU or accelerators to meet time-sensitive user needs.

\noindent{\bf Algorithmic Attacks and Defenses:} 
Existing ML attacks primarily target vulnerabilities in the algorithm itself, regardless of the deployment environment. Such attacks include membership inference, model extraction, and data poisoning, among others and are classified as \emph{algorithmic attacks} in our document.
Membership inference attacks~\cite{shokri2017membership} aim to determine if specific data were part of a model's training set, risking exposure of sensitive private information to the public. Techniques like machine unlearning~\cite{bourtoule2021machine} can defend against these attacks.
Model extraction attacks~\cite{tramer2016stealing, milli2019model} seek to replicate a model’s architecture and parameters, which are valuable intellectual properties requiring significant computational resources. Thus, it is important to prevent model extraction~\cite{juuti2019prada}.
Data poisoning attacks~\cite{tian2022comprehensive} involve injecting corrupted data into the training set, either reducing overall model accuracy or targeting specific predictions. Data augmentation~\cite{borgnia2021strong} is a defense strategy against such attacks.
These attacks rely solely on model inputs and outputs during training or inference, without considering side-channel data tied to the deployment environment. They are distinct from system attacks we categorize in this work.


\section{Threat Model Categorization}
\label{sec:threat_model}

The diversity of machine learning deployments demands a broad and adaptable threat model. Software and hardware defenses often depend on different threat models, complicating the evaluation of an attack’s relevance in a given context. Moreover, studies vary in their focus on secret assets and their assumptions about trusted entities.

To address this, we first align the AI/ML attack landscape with the Mitre Att\&ck framework~\cite{strom2018mitre}, providing a structured approach to understand threat models. We then introduce various assets within a compound AI system, and identify the trusted entities in these deployments.

\subsection{MITRE ATT\&CK Framework}
\label{subsec:tm_mitre}
MITRE ATT\&CK~\cite{strom2018mitre} is a knowledge base that models cyber adversary behavior using various real-world observations. This framework reflects the various phases of an adversary’s attack lifecycle and the platforms they are known to target.
We align the system and hardware attacks on AI/ML systems to the MITRE framework to enable threat model development and explore defense methodologies in AI applications. ~\Cref{fig:mitre_tm} translates various attacks seen across literature and Common Vulnerability and Exposure(CVE) databases as as techniques across different categories from the framework: 
\begin{enumerate}[leftmargin=*]
    \item \textbf{Reconnaissance and Initial Access} lists the techniques used by an attacker to identify and locate a victim model or target device to mount an attack. For example, poisoning of training data or exploiting co-location patterns~\cite{zhao2024everywhere} on cloud allows attackers to target specific applications or models.
    \item \textbf{Resource Development} lists down methodologies an attacker utilizes to prepare for an exploit. These might include employing shared memory regions with victim~\cite{vila2019theory} and exploiting access control violations.
    \item \textbf{Privilege Escalation and Execution} enumerates methods for adversaries to realize an attack. These methods can be software or hardware contention~\cite{flushReload,primeProbe}, and passing illegal prompts that lead to system errors or confused deputy~\cite{confusedpilot} issues.
    \item \textbf{Collection and Exfiltration} identifies the types of metrics or data that an attack might expose, revealing sensitive information (e.g., performance counters, power traces, execution time) or causing service-level disruptions (e.g., misclassification, Denial of Service).
    \item \textbf{Impact} lists down the confidentiality, integrity and availability aspects of an attack result as seen in the ``Collection and Exfiltration" techniques.
\end{enumerate}

\subsection{Threat Models}
We divide the attacker's scope into categories that we define here as four threat models. In \cref{fig:mitre_tm}, we enumerate techniques in increasing order of threat model severity. Darker colored rows indicate a higher privilege adversary, with attacks having a higher severity impact.

\noindent{\bf Remote software access}
describes a threat model where an adversary only has remote access to a compound AI system. This is typical for cloud-hosted applications providing API access to AI applications. Here, the attacker is least-privileged since she can only influence execution indirectly (e.g. resource contention) or leak data via prompt injection.

\noindent{\bf Privileged software access}
involves an adversary with elevated software control, such as root access to an OS, hypervisors, or firmware. This includes untrusted system administrators in a public cloud environment and users exploiting privilege escalation vulnerabilities in AI pipelines. An adversary can monitor execution via system traces, device drivers, or disrupt operations by manipulating schedulers.

\noindent{\bf Digital hardware access}
includes adversaries that can issue system commands for observing hardware metrics (e.g., performance counters) or launch covert-channel attacks to leak information like model weights and layer architecture. This threat model is common in public cloud environments where attackers co-locate malicious workloads with AI pipelines.

\noindent{\bf Physical hardware access}
represents the most privileged adversary, with full physical access to hardware running AI inference. This is common in edge ML deployments where devices running sensitive models are controlled by untrusted owners, allowing potent attacks like cold-boot, memory dumps, or access to debug ports.

\subsection{Secret Assets}
\label{subsec:assets}

Secret assets refer to confidential information that attackers aim to exploit. Each component of a compound AI system possesses distinct secret assets, as outlined below:

\noindent{\bf Model training data} 
A compound AI system is composed of multiple large language models (LLMs). 
These models are trained on extensive datasets, some of which include private or confidential information.
Furthermore, fine-tuning these models is often carried out using proprietary data.
Prior work~\cite{carlini} leaked confidential information including user ssh key, user name, credit card info, address etc. from training data.
Membership inference attacks determine whether specific information was included in the training data.
Lastly, certain attacks analyze the distribution of the training data and use it to compromise the inference process.


Data poisoning attacks can either skew a model's decisions or diminish its overall effectiveness. Such attacks are more challenging to be detected in multi-modal training due to adversarial noise -- highlighting the importance of integrity in training data.

\noindent{\bf AI model IPs}
LLMs contain billions of parameters and takes significant training resources, making them lucrative attack targets.
Attackers have tried to violate confidentiality of ML models by extracting the weights, architecture and hyperparameters. 
The model architecture can be private for proprietary foundation models and confidential adapters for fine-tuned models. 
Similarly, model hyperparameters including learning rate, and others are lucrative for attackers.
Additionally, the semantic relationships between tokens in an embedding model are confidential, as attackers can exploit this information to carry out membership inference attacks.
The integrity of the model data is equally important. 
Altering model hyperparameters can degrade model accuracy, while tampering with model weights may lead to mis-classifications or the spread of misinformation.

\noindent{\bf Knowledge database}
The LLM models leverage recent knowledge from vector databases, which are constantly being populated with new information. 
The integrity of the vector database is critical for generating relevant and correct information. 
Similarly, the vector DBs should not be susceptible to availability attacks like denial-of-service, which can generate stale response. 

\noindent{\bf Inference data}
User queries to a compound AI system can contain sensitive information like medical records, personally identifiable information(PIIs) or financial documents. For such private data a user would like to ensure confidentiality from the software, model owners and other tenants of the compound AI application.

\subsection{Trust Entities}
\label{subsec:trust_entities}
The relationship between different trust entities plays key role in defining a threat model. 
The entities include:

\noindent{\bf Hardware manufacturer}
has access to the hardware design and the manufacturing supply chain.
Hardware Trojans introduce backdoors that can be used to exploit CIA guarantees for all assets. 
Since, a lot of ML systems are designed on FPGAs, the role of hardware manufacturer includes the FPGA manufacturer and the hardware design bitstream owner.

\noindent{\bf Platform owner}
refers to a cloud-service provider or an admin of the hosted hardware. A platform owner multiplexes different model run on the same hardware to improve the resource utilization of the cluster. A platform owner or a hypervisor admin can have access to privileged software like devise drivers or have physical access to hardware to mount various snooping/physical attacks.

\noindent{\bf Training data owner}
refers to owner of private data that is used in training an ML model. Although modern LLMs are typically trained on a large corpus of public data, many expert models are then fine-tuned by training on private and sensitive data (e.g. medical images, financial documents). Algorithmic attacks~\cite{shokri2017membership} have been shown to divulge training data using membership inference attacks.

\noindent{\bf Model owner}
refers to an owner of an ML model where the model weights, layer architecture and the hyperparameters could be private and sensitive information. Model owners are susceptible to information leakage via algorithmic attacks~\cite{milli2019model,tramer2016stealing} and from distrusted co-located tenants via side-channels~\cite{hua2018reverse,cacheTelepathy}.

\noindent{\bf Inference data owner}
is a client side user who provides data for inference to the AI/ML application. Similar to training data, these inputs can be sensitive and contain PIIs which an user would want to keep confidential from other users and tenants on the platform.

\noindent{\bf Software developers}
refer to owners of third-party services and framework components that are associated with an end-to-end ML/AI application. Bugs and vulnerabilities can enable attackers to gain unrestricted access to the runtime service or the ability to mount man-in-the-middle attacks to leak sensitive information.

\section{Software vulnerabilities and defenses}
\label{sec:sw_att_def}

Design of compound AI systems involve a \textit{development} stage followed by a \textit{deployment} stage that involve a wide variety of software frameworks, packages and libraries. The \textit{development stage} focuses on model training, knowledge integration and designing of LLM pipelines that are built using frameworks like Langchain~\cite{Langchain} and HuggingFace~\cite{HuggingFace} and model compilation packages(Tensorflow~\cite{Tensorflow}, Pytorch~\cite{Pytorch}). Vulnerabilities like trojans, misconfigured access control of databases, and buggy libraries and drivers can lead to privilege escalation and initial access into systems that are leveraged to mount sophisticated attacks. Verification of large codebases in compound AI systems are unfeasible because of the computational difficulty involved in with extensive repositories and models. Additionally the challenge of accurately modeling the interactions between libraries and continuous evolution of such deployments make it impractical to rely only on software verifications.

The \textit{deployment stage} relies on various data storage (Apache Spark~\cite{Spark}, Hadoop~\cite{Hadoop} and Snowflake~\cite{Snowflake}) and model orchestration (BentoML~\cite{BentoML}, Kubernetes~\cite{Kubernates}) to deploy AI pipelines across heterogeneous platforms that utilize device libraries and packages (Pyyaml~\cite{Pyyaml}, CuDNN~\cite{Cudnn}) to provide optimized executions. Vulnerabilities and lack of safeguards across the software stack and device drivers can leak to black-box attacks like prompt injection and membership inference, or model tampering and model fingerprinting attacks that violate the confidentiality, integrity and availability of compound AI systems.

\begin{table*}
    \centering
    \caption{\textbf{List of software components with vulnerabilities in compound AI frameworks, packages and underlying libraries.\sqTypeB requires a privileged attacker while \sqTypeC refers to DoS attacks.}}
    \vspace{-1em}
    \resizebox{1\linewidth}{!}{%
    \setlength{\tabcolsep}{0.7pt}
    \begin{tabular}{|c|c|c|c|c|c|}
        \hline
        \typeD{\textbf{\parbox[c]{2cm}{}}} &
        \typeD{\textbf{\parbox[c]{3.8cm}{\centering Data Confidentiality}}} &
        \typeD{\textbf{\parbox[c]{3.8cm}{\centering Data Integrity}}} &
        \typeD{\textbf{\parbox[c]{3.8cm}{\centering Crash/DoS}}} & 
        \typeD{\textbf{\parbox[c]{3.8cm}{\centering Code Execution}}} &
        \typeD{\textbf{\parbox[c]{3.8cm}{\centering Privilege Escalation}}}
        \\
        \hline
        \hline
        &&&&&\\[-0.8em]
        \multirow{6}{*}{\rotatebox[origin=c]{90}{\parbox[c]{2cm}{\centering \textbf{ Frameworks}}}} &
        \typeA{\parbox[c]{3.8cm}{\centering \textbf{Langchain}   - SQL Injection     \\ CVE-2023-36189}} &
        \typeB{\parbox[c]{3.8cm}{\centering \textbf{ChatGPT}     - Integrity         \\ CVE-2024-40594}} &
        \typeC{\parbox[c]{3.8cm}{\centering \textbf{NLTK}        - ReDoS             \\ CVE-2021-43854}} &
        \typeA{\parbox[c]{3.8cm}{\centering \textbf{Haystack}    - ACE               \\ CVE-2024-41950}} &
        \typeA{\parbox[c]{3.8cm}{\centering \textbf{Langchain}   - SSRF              \\ CVE-2024-3095 }} \\[5pt]
        &&&&&\\[-0.8em]
         &
        \typeA{\parbox[c]{3.8cm}{\centering \textbf{LLama}        - OOB Read         \\ CVE-2024-42478 }} &
        \typeA{\parbox[c]{3.8cm}{\centering \textbf{LLama}        - OOB Write        \\ CVE-2024-42479 }} &
        \typeA{\parbox[c]{3.8cm}{\centering \textbf{LLama}        - Heap Ovf.        \\ CVE-2024-42479 }} &
        \typeA{\parbox[c]{3.8cm}{\centering \textbf{HuggingFace}  - RCE              \\ CVE-2024-3568  }} &
        \typeA{\parbox[c]{3.8cm}{\centering \textbf{Langflow}     - ACL Violation    \\ CVE-2024-7297  }} \\[5pt]
        &&&&&\\[-0.8em]
        &
        \typeA{\parbox[c]{3.8cm}{\centering \textbf{HuggingFace}  - Access Control   \\ CVE-2023-2800  }} &
        \typeA{\parbox[c]{3.8cm}{\centering \textbf{Rasa}         - Directory Trvsl  \\ CVE-2021-42556 }} &
        \typeC{\parbox[c]{3.8cm}{\centering \textbf{Mxnet}        - Resource Hog     \\ CVE-2022-24294 }} &
        \typeA{\parbox[c]{3.8cm}{\centering \textbf{LLamaindex}   - Bug/RCE          \\ CVE-2024-3271  }} &
        \typeA{\parbox[c]{3.8cm}{\centering \textbf{Ollama}       - ACL Violation    \\ CVE-2024-28224 }} \\[5pt]
        &&&&&\\[-0.8em]
        \hline
        &&&&&\\[-0.8em]
        \multirow{6}{*}{\rotatebox[origin=c]{90}{\parbox[c]{2cm}{\centering \textbf{Packages}}}} &
        \typeA{\parbox[c]{3.8cm}{\centering \textbf{Pytorch}     - OOB Read         \\ CVE-2024-31584 }} &
        \typeA{\parbox[c]{3.8cm}{\centering \textbf{MLFlow}      - ACL Violation    \\ CVE-2024-4263  }} &
        \typeC{\parbox[c]{3.8cm}{\centering \textbf{Snowflake}   - Arbit. input     \\ CVE-2022-42965 }} &
        \typeA{\parbox[c]{3.8cm}{\centering \textbf{Pytorch}     - Cmd Injection    \\ CVE-2022-0845  }} &
        \typeA{\parbox[c]{3.8cm}{\centering \textbf{Pytorch}     - API Bug          \\ CVE-2024-5480  }} \\[5pt]
        &&&&&\\[-0.8em]
         &
        \typeB{\parbox[c]{3.8cm}{\centering \textbf{TensorFlow}  - Use-after-free   \\ CVE-2021-41220 }} &
        \typeA{\parbox[c]{3.8cm}{\centering \textbf{TensorFlow}  - OOB Write        \\ CVE-2022-41894 }} &
        \typeA{\parbox[c]{3.8cm}{\centering \textbf{TensorFlow}  - FP Exception     \\ CVE-2023-27579 }} &
        \typeA{\parbox[c]{3.8cm}{\centering \textbf{Snowflake}   - Cmd. Injection   \\ CVE-2023-34230 }} &
        \typeA{\parbox[c]{3.8cm}{\centering \textbf{Hadoop}      - ACL Violation    \\ CVE-2023-26031 }} \\[5pt]
        &&&&&\\[-0.8em]
         &
        \typeB{\parbox[c]{3.8cm}{\centering \textbf{Pytorch}     - Malicious Pkg.   \\ CVE-2022-30877 }} &
        \typeB{\parbox[c]{3.8cm}{\centering \textbf{Spark}       - Cleartext Store  \\ CVE-2019-10099 }} &
        \typeC{\parbox[c]{3.8cm}{\centering \textbf{Kubernetes}  - Bug              \\ CVE-2022-3172  }} &
        \typeA{\parbox[c]{3.8cm}{\centering \textbf{BentoML}     - RCE              \\ CVE-2024-2912  }} &
        \typeA{\parbox[c]{3.8cm}{\centering \textbf{Kubernetes}  - API Bug          \\ CVE-2023-1260  }} \\[5pt]
        \hline
        &&&&&\\[-0.8em]
        
        \multirow{6}{*}{\rotatebox[origin=c]{90}{\parbox[c]{2cm}{\centering \textbf{Libraries}}}} &
        \typeB{\parbox[c]{3.8cm}{\centering \textbf{OpenCL}      - Use-after-free   \\ CVE-2023-4969  }} &
        \typeA{\parbox[c]{3.8cm}{\centering \textbf{Py zlib}     - Mem Corrupt      \\ CVE-2018-25032 }} &
        \typeA{\parbox[c]{3.8cm}{\centering \textbf{Scipy}       - Heap Ovf.        \\ CVE-2022-48560 }} &
        \typeA{\parbox[c]{3.8cm}{\centering \textbf{Pyyaml}      - RCE              \\ CVE-2021-4118  }} &
        \typeA{\parbox[c]{3.8cm}{\centering \textbf{OneAPI}      - Library Bug      \\ CVE-2024-21766 }} \\[5pt]
        &&&&&\\[-0.8em]
         &
        \typeA{\parbox[c]{3.8cm}{\centering \textbf{CuDNN}       - Use-after-free   \\ CVE-2021-37652 }} &
        \typeA{\parbox[c]{3.8cm}{\centering \textbf{SQLite}      - OOB Write        \\ CVE-2020-35527 }} &
        \typeB{\parbox[c]{3.8cm}{\centering \textbf{Mlnx OS}     - IPfilter config  \\ CVE-2024-0101  }} &
        \typeA{\parbox[c]{3.8cm}{\centering \textbf{Redis}       - Integer Ovf.     \\ CVE-2023-41056 }} &
        \typeA{\parbox[c]{3.8cm}{\centering \textbf{Py Urllib}   - Inp. Validation  \\ CVE-2023-24329 }} \\[5pt]
        &&&&&\\[-0.8em]
         &
        \typeB{\parbox[c]{3.8cm}{\centering \textbf{Cuda}        - OOB Read         \\ CVE-2023-25513 }} &
        \typeB{\parbox[c]{3.8cm}{\centering \textbf{vGPU Drv}    - OOB Write        \\ CVE-2023-31035 }} &
        \typeB{\parbox[c]{3.8cm}{\centering \textbf{Firmware}    - Config Error     \\ CVE-2023-31035 }} &
        \typeB{\parbox[c]{3.8cm}{\centering \textbf{vGPU Drv}    - OOB Write        \\ CVE-2023-31035 }} &
        \typeA{\parbox[c]{3.8cm}{\centering \textbf{Torchserve}  - Directory Trvsl  \\ CVE-2023-48299 }} \\[5pt]
        \hline
    \end{tabular}
    }
    \label{tbl:sw_att_list}
\end{table*}

\subsection{Attacks}
\label{subsec:sw_att}

We searched the CVE database to report several vulnerabilities across different components (Frameworks, Packages and Libraries) of the compound AI stack in~\cref{tbl:sw_att_list}.
The columns categorize the attacks based on the attackers' motive, which includes violating data confidentiality, integrity, and availability (C,I,A) of the system. 
The last two columns enable the attacker to either run arbitrary code in the victim machine, or performing privilege escalation. 
These attacks enable the attacker to collect victim information including access to system logs, scheduling processes, linking malicious libraries, or introduce ransomware.
These capabilities often lead to model hijack and tampering attacks. 
We describe each of the attack types in detail.

\noindent{\bf Data Confidentiality}
An attacker can leak asset classes including model weights and hyperparameters (layer architecture, learning coefficient, temperature etc.), input query, training data and documents in the vector database.
The common weaknesses leading to data confidentiality violation include 
\textbf{(1)} Buffer out-of-bound reads (OOB Read);
\textbf{(2)} Use-after-free memory errors;
\textbf{(3)} Backdoor introduced by malicious packages; and
\textbf{(4)} System bugs leading to insufficient access control.
The first column in~\cref{tbl:sw_att_list} lists several CVEs leading to data confidentiality violation, with the green boxes needing privileged system access.
The main causes of information leakage are \textit{insufficient access control} leading to active or stale data leakage from registers and memory.
\textit{Software bugs or malicious packages} trigger memory safety errors and 
The majority of compound AI software is written in Python and C++ which are not \textit{type safe} languages, and hence lacking compiler protections.

\noindent{\bf Data Integrity}
Data integrity is essential to prevent mis-classification in edge deployments and the spread of misinformation in complex RAG systems. 
The common weaknesses include 
\textbf{(1)} Buffer out-of-bound writes (OOB Writes) by an attacker into victim memory.
\textbf{(2)} Access control violations enable an unprivileged attacker to gain access to victim data while cleartext writes in databases enable admin to change document contents.
\textbf{(3)} Directory traversal enables attackers to gain write access to model files while memory corruption on compression libraries (Python zlib) either tampering or deleting sensitive data.

\noindent{\bf Crash/Denial-of-Service}
System or platform availability can be restricted with system crashes or denial-of-service (DoS) attacks as shown in the third column of~\cref{tbl:sw_att_list}.
The system crashes are triggered with unexpected behavior including: 
\textbf{(1)} Heap overflow attacks leading to kernel panic; or
\textbf{(2)} System exceptions specifically in floating point units(FPE) or integer overflows leading to system crash.
The DoS attacks include 
\textbf{(1)} An attacker taking over the entire system or model resource preventing other tenants from accessing the model;
\textbf{(2)} An attacker can also target the Kubernetes orchestration to prevent other tenants from accessing a compute node. 
\textbf{(3)} Creating DoS by engaging the system in a complex regular expression (ReDoS) is shown in NLTK framework. 
System availability is important in mission critical applications (autonomous vehicles, industry quality-check installations)  or real-time usecases (Chatbots).

\noindent{\bf Code execution}
Several compound AI systems are deployed in a large cloud infrastructure. 
A code execution vulnerability can be exploited to reveal system configuration or 
identity. Moreover, arbitrary code can interfere with system robustness or trigger inaccessible tools inside the AI pipeline. 
\textbf{(1)} Arbitrary code execution (ACE) enables attackers to convert non-executable memory regions into malicious scripts. 
\textbf{(2)} Remote code execution (RCE) attacks perform the same on a remote system and is even more catastrophic.
\textbf{(3)} Command injection attacks are able to query secret datatypes by inserting a command. For instance, victim database entries are deleted by inserting 
\texttt{delete()} command in a Snowflake database. 
\textbf{(4)} The OOB write vulnerability in the vGPU driver~\cite{cveL2.3} flips device configuration enabling an attacker to perform unauthorized code execution, leading to device hijack.

\noindent{\bf Privilege escalation}
The green boxes in~\cref{tbl:sw_att_list} requires attackers to have a privileges system access. 
However, a non-privileged attacker can mount privilege escalation attack to become the root user of the system. 
The privilege escalation vulnerabilities include:
\textbf{(1)} Access control (ACL) violations in software frameworks and packages pose significant risks. Many of these frameworks, packages, or libraries include kernel modules, allowing attackers to exploit ACL violations or API bugs to gain privileged access.
\textbf{(2)} Directory traversal or the lack of input validation is exploited by attackers with blackbox access to access privileged data structures. 
\textbf{(3)} Server side request forgery (SSRF) is used by a remote attacker to gain 
unauthorized system access and we found one such CVE in Langchain~\cite{cveF5.1}.

\subsection{Defenses}
\label{subsec:sw_def}

The complex software stack in a compound AI system is riddled with vulnerabilities at all levels as discussed in~\cref{subsec:sw_att}.
While many of the frameworks, libraries and packages might have individual safety nets, it is clearly insufficient to protect against different vulnerability classes.
Moreover, the active development and the sheer size of the codebase makes it infeasible for formal verification techniques~\cite{FV} or even high-coverage fuzzing methods~\cite{fuzz}. 
A key observation though is that, some vulnerabilities like buffer overflow, use-after-free, and memory errors etc. can be thwarted with existing defenses. 
However, certain other attacks like victim co-location, input validation or installation of malicious package catalyze certain algorithmic or hardware attacks.
In this section, we propose defenses against direct software vulnerabilities as well as mandate certain software practices to prevent leveraging system software in performing algorithmic and hardware attacks.

\noindent{\bf Software supply-chain defenses}
Protecting the software supply-chain is a first-line defense against the listed vulnerabilities.
The Compound AI software stack incorporates numerous dependencies sourced from a wide range of vendors.
Prior work~\cite{sw_supply_chain} proposes user credential and role validation to prevent package tampering.

The second challenge is the authentication of package registry. The attacker registers several malicious packages with similar names to trick the victim. Prior work~\cite{ladisa2023sok} employs administrative safeguards as well as remove unused dependencies to protect against these attacks. Amalfi~\cite{amalfi} uses ML classifiers to test package reproducibility from source code to automatically detect malicious packages.

\noindent{\bf Access control policies}
Declaration of variables with an appropriate access modifier such as \texttt{private}, \texttt{public} and \texttt{protected} as provided by languages like Python, C++ and Java is necessary to avoid secret data leakage and tampering due to improper access control policies.
Secret variables should always be declared as \texttt{private} to prevent access from different class methods. 
Code review and information flow tracking compiler passes~\cite{sw_ift1,raksha,sw_ift2} can help prevent secret data leakage from training datasets, model parameters and knowledge vector database.
Fine-grained identity and access management policies (IAM) should also be validated~\cite{wutschitz2023rethinkingprivacymachinelearning} for datasets, model and other resources.

Input queries and retrieved context should undergo input validation to prevent privilege escalation attacks. 
Input validation should verify token lengths, detect malicious regular expressions, and sanitize retrieval commands. 
Exhaustive testing of the trained model eliminates unknown behavior when presented with crafted inputs. 
Safeguards should be placed at different compound AI components both on the client and server side to prevent access control attacks.

\noindent{\bf Memory safety}
We discovered several CVEs for different types of buffer overflow attacks, dangling pointers and use-after-free errors.
Illegal memory accesses lead to confidentiality, integrity and availability concerns. 
While defenses are proposed for these well-known problems, the deployment scale and the active development leads to these errors. 
There is an urgent need to shift the software backend from C,C++ to memory safe languages like Rust.
The software community is developing critical packages~\cite{rust_ai} and linux OS~\cite{rusty} in Rust language to ensure memory safety. 
Running compound AI systems in a sandbox environment~\cite{erim,securecells,rlbox,hfi} reduces the risk of privilege escalation and DoS attacks, which are particularly catastrophic in cloud deployments.
Existing memory safety protections including address-space-linear randomization (ASLR), stack canaries and use-after-free protections should be enabled for all production systems.
Memory tagging techniques like CC, Cheri~\cite{cheri} and Morpheus~\cite{morpheus} isolate data from different security domains and are implemented in latest Intel and arm processors to deter memory safety errors.

\noindent{\bf Node resiliency and check-pointing}
Training of foundational models take place in vast distributed systems with hundreds of GPUs and terabytes of storage. 
Orchestration frameworks are used for compute scheduling and efficient data movement. 
This frameworks should be fault tolerant and be able to deal with node crashes. 
For instance, orchestration frameworks can migrate computations away from specific nodes~\cite{kubernates_resilience} based on system's health to prevent large scale infrastructure failures. 
The ML model should be check-pointed at regular and frequent intervals to ensure minimal loss during a large scale failure. 
Moreover, efficient log collection and forensic analysis is required for root causing the failure.
Omnilog~\cite{omnilog} collects Linux audit logs efficiently and ensure log integrity after system compromise. 
Recent works have employed machine learning techniques to find malicious attack signatures for faster attack detection.

\subsection{Key Takeaways}
\label{subsec:sw_kt}

\textbf{\textit{Takeaway 1: Safeguarding the software supply chain:}}
The compound AI software stack contains a large number of software packages from multiple vendors. 
This creates a distributed attack surface enabling an attacker to inject a malicious package or software backdoors to extract secret data. 
Software supply chain defenses should validate the developer, leverage packages from well-known sources to ensure secret data confidentiality and integrity protection.

\textbf{\textit{Takeaway 2: Design with type-safe languages:}}
A majority of data leakage and privilege escalation attacks are a result of improper access control policies and memory safety errors. 
Memory- and type-safe languages like rust should be used to prevent these attacks.

\textbf{\textit{Takeaway 3: Software defenses for heterogeneous platforms:}}
Many software defenses (e.g. ASLR, stack canary) found on CPU environments are crucial on platforms like GPUs, ML accelerators~\cite{tpu} and even DPUs~\cite{dpu} since rise in AI deployments have exposed these platforms to threat models similar to CPU systems. 

\textbf{\textit{Takeaway 4: System wide software health monitors}}
Small errors in multiple software components can cause a catastrophic system failure. 
Hence, system-wide software monitors should be deployed in the orchestration layer to monitor the health of each deployed node. 
Resource hogging or access control violations should be detected proactively to minimize system impact. 

\section{Hardware vulnerabilities and defenses}
\label{sec:hw_att_def}

\begin{table*}[t]
    \centering
    \resizebox{0.6\linewidth}{!}{%
    \begin{tabular}{lllll}
        \hline
        \textbf{Component} &
        \textbf{Attack Type} &
        \textbf{Channel} &
        \textbf{Asset} &
        \textbf{Breach}
        \\
        \hline
        \multirow{7}{*}{\shortstack{Memory}} &
        Boot attacks~\cite{coldboot,warmboot} &
        Direct &
        $M_v,I$ & C\\
        \cline{2-5}
        & Bank Conflict~\cite{drama,dramaqueen} &
         \multirow{2}{*}{Timing} &
        $M_a$ & C\\
        & NVleak~\cite{nvleak} &
         &
        $M_a$ & C\\
        \cline{2-5}
        & Rowhammer~\cite{deephammer,oneflip,bitflip} &
        \multirow{2}{*}{Bitflip} &
        $M_v, I$ & I\\
        & Rambleed~\cite{rambleed,deepsteal} & &
        $M_v, I, T$ & C\\
        \cline{2-5}
        & PiM/PnM attacks~\cite{imc_rram_power,powergan,impact} &
        Power &
        $M_v, I$ & C\\
        \cline{2-5}
       & Laser attacks~\cite{laserFltInj,fault_inj1,fault_inj2} &
        Injection &
        $M_v, I$ & I\\
        \hline

        \multirow{9}{*}{\shortstack{Interconnect}} &
        Bus hijack~\cite{pcileech,thunderclap,chipclone} &
        \multirow{3}{*}{Direct} &
        $M_v,I, T$ & C,I\\
        & Access pattern~\cite{reverseCNN,band_util,band_util_2,hmtt,hermes} &
        &
        $M_a$ & C\\    
        & Perf. Counters~\cite{deepsniffer,GPU_cupti,leakydnn} &
        &
        $M_a$ & C\\  
        \cline{2-5}
        & Sparsity~\cite{huffduff,sparsityTime} &
        \multirow{3}{*}{Timing} &
        $M_v$ & C\\  
        & Contention~\cite{nvlink_attack,invisiprobe} &
        &
        $M_a$ & C\\  
        & Fingerprinting~\cite{deepsniffer, memAccess, layerExtract} &
        &
        $M_a$ & C\\    
        \cline{2-5}
        & HW Trojan~\cite{bus_snoop} &
        Bitflip &
        $M_v,I,T$ & I \\
        \cline{2-5}
        & TDC with RO~\cite{tdc} &
        Power &
        $M_a$ & C \\
        \cline{2-5}   
        & Voltage Virus~\cite{majumdar2024voltage} &
        Injection &
        $M_v, I$ & I \\
        \hline 
        \multirow{6}{*}{\shortstack{Compute}} &
        Spad use-after-free~\cite{banerjee2020sesame,mindctrl} &
        Direct &
        $M_v,I, T$ & C,I\\
        \cline{2-5}
        & Cache Attacks~\cite{cacheTelepathy,deeprecon,ganred,spyGpuBoX} &
        \multirow{2}{*}{Timing} &
        $M_v, T,I$ & I\\  
        & Exec. Time~\cite{csi_nn,compute_sema,barracuda,fpmt} &
         &
        $M_a, M_v$ & I\\  
        \cline{2-5}
        & HW Trojan~\cite{bias_buffer} &
        Bitflip &
        $M_v, T,I$ & I\\  
        \cline{2-5}
        & 
        Fingerprinting~\cite{opendnn, power_survey}&
        Power &
        $M_a$ & C\\  
        \cline{2-5}   
        &
        Laser~\cite{laser_sram} &
        Injection &
        $M_v,T,I$ & I\\ 
        \hline
    \end{tabular}
    }
    \caption{\textbf{List of attacks in different hardware components. Asset classes include model weights ($M_v$), model architecture ($M_a$), training data ($T$) and private query ($I$). }}
    \vspace{-2em}
    \label{tbl:hw_att_list}
\end{table*}

Beyond the algorithmic and software attacks, an attacker can mount direct, side-channel and bit-flip attacks in hardware.
These attacks can be performed both during training and query inference in a compound AI system. 
We define \textit{direct attacks} as those in which the attacker gains unauthorized access to read or manipulate confidential data directly from storage, memory, interconnect or on-device buffers.
A \textit{timing attack} is a side-channel attack where the attacker extracts victim secret data from execution timing variation.
An attacker can use hardware performance monitors, or high-resolutions timers to extract fine-grained victim execution timing.
A \textit{power attack} extracts power signatures to infer victim execution.
\textit{Bitflip} and \textit{injection} attack tampers secret data.
A bitflip attack is non-invasive, while an injection attack requires hardware physical access.

In this section, we will first enlist different attacks performed on ML systems to extract different assets including training data privacy, model parameters and hyper-parameters, and user inference input. 
Next, we will discuss hardware defense mechanisms deployed to protect confidentiality and integrity of ML execution, with a final key takeaway section summarizing the main design trends and their mapping to the attack vectors.

\subsection{Attacks}
\label{subsec:hw_att}
We categorize existing hardware attacks into different attack categories. 
It includes attacks on heterogeneous hardware including CPUs, GPUs and ML accelerators.
We categorize the hardware components into \textit{memory}, \textit{interconnect} and \textit{compute} components as shown in ~\cref{tbl:hw_att_list}. 
\textit{Memory} attacks encapsulate data leakage and tampering in main memory. 
\textit{Interconnect} attacks cover leakage from memory bus and I/O interconnect.
\textit{Compute} attacks encompass micro-architectural attack on on-chip memory, buffers and processing units. 
The attacker can either perform hardware attacks remotely (digital attacks) or require physical system access (physical attacks).

Digital attacks include timing side-channels and resource utilization attacks while \textit{physical attacks} include power and fault injection attacks that require device physical access to snoop signals or tamper with execution.

\noindent{\bf Memory and storage attacks}
Memories are used to store proprietary ML models and secret user data.
CPUs, GPUs and accelerators are typically connected to DDR rams, SD rams or an HBM storage.
Attackers target data stored in these memory components in order to compromise its confidentiality and integrity.
\textit{Direct} attacks dump secret data stored as plaintext from memory. 
Boot attacks exploit the cell retention of volatile memories across reboots to leak secret data, violating confidentiality of model parameters ($M_v$), secret user inputs (I) and training data (T).
\textit{Timing} attacks retrieve model architecture ($M_a$) by recording the timing difference between row buffer hits and misses. 
NVLeak~\cite{nvleak} attack leaks secret vector database knowledge stored in persistent storage. 
\textit{Bitflip} attacks like Rowhammer~\cite{mutlu2019rowhammer} and Rowpress~\cite{luo2023rowpress} can flip model parameters to reduce model performance. Many large models are resilient to bitflip attacks, especially for early layer bitflips. 
However, these attacks can be used to tamper user input queries leading to either mis-classification or model hallucination.
Rambleed attacks use rowhammer to leak secret data from adjacent rows. This can also be used to infer user queries, as well as targeted training data and model parameters. 
Despite rowhammer mitigations~\cite{saxena2022aqua} in latest generations of memory, recent work~\cite{luo2023rowpress} demonstrated rowhammer in HBM memories widely used in latest GPUs.
\textit{Power} attacks are demonstrated on processing-in-memory (PiM) and processing-near-memory (PnM) devices, which are gaining popularity due to data transfer minimization.
Specifically RRAM PiMs emit power signatures based on layer architecture. 
PnM access pattern is also used to infer confidential DNA sequence input. 
Finally, laser is used inject faults in memory, which can poison training data samples or tamper model parameters during inference. 

\noindent{\bf Interconnect attacks}
The next attack surface is on the memory bus and I/O interconnect. 
This is widely used to data transfer between the compute units like CPUs, GPUs and accelerators and memory and storage blocks like DRAM and nvme drives. 
\textit{Direct} attacks include bus hijack by either connecting a malicious device over the PCIe or using a bus monitor. 
Bus hijack attacks can be used to either read secret data during transfer or perform a man-in-the-middle attack to tamper input queries or ML models. 
The interconnect is widely used across heterogeneous compute units in a compound AI system, making it lucrative for attackers. 
Hijacking the interconnect enables the attacker to control the decisions taken by the LLM agent and emit malicious response.
Demand access pattern extracted from memory bus is also used to infer model layer architecture. 
Finally, several performance counters reveal read/write data volumes to fingerprint service usage or layer connections in a ML model. 
While, performance counters are mostly disabled in production systems, CSPs can use them for analytics, leading to application and service fingerprinting attacks.
\textit{Timing} attacks include leakage from data-driven optimizations like model sparsity. Prior work~\cite{sparsityTime} uses AXI bus monitors to reveal sparsity ratio from data volume. 
Bus contention~\cite{band_util} is used to snoop network traffic, revealing model architecture. 
Finally, bus utilization is used to fingerprint model layer dimensions. 
Hardware Trojan tampers model parameters to misclassify ML inference. 
Physical side-channels include ring oscillators to snoop power consumption, which is used as a proxy to infer model dimensions. 
These attacks reveal model architecture, which is then used to perform membership inference or model extraction attacks with white-box assumption. 
Voltage viruses~\cite{majumdar2024voltage} can be used to tamper with model inference in multi-tenant FPGA deployments.

\noindent{\bf Compute attacks}
On-chip microarchitectural sharing of multiple tenants expose several attack vectors.
Sesame~\cite{banerjee2020sesame} observed several attacks for resource sharing in multi-tenant accelerators. 
These include sharing of compute blocks and use-after-free attacks in a spatially shared scratchpad.
\textit{Timing} attacks include CPU and GPU caches. 
Cachetelepathy~\cite{cacheTelepathy} leveraged prime+probe and flush+reload attacks to leak ML model architecture. Similar attacks were demonstrated in GPU caches as well. 
Moreover, compute execution time is inferred from EM emission~\cite{csi_nn} and from floating-point unit~\cite{fpmt} and used to fingerprint different layer types and dimensions. 
These works also use novel algorithms to efficiently reverse engineer the model architecture. 
\textit{Bitflip} attacks are used to tamper the bias buffer~\cite{bias_buffer}, which leads to model mis-classification. 
Misclassification in the LLM agent can lead to incorrect response and trigger incorrect tools for query inference. 
Prior work~\cite{power_survey} used power channel and magnetic signals to fingerprint layers by targeting the GEMM module. 
Finally, fault injection in control queues can violate dependency queues in a DAE design. This triggers model computation before the data is ready in scratchpad leading to incorrect model response.

\subsection{Defenses}
\label{subsec:hw_def}

Prior works has proposed several defense mechanisms to protect secret data from the attacks discussed in ~\cref{subsec:hw_att}.
Some of the defenses target specific micro-architectural components like caches, scratchpad and compute units, while others strive to design isolated execution systems.

\noindent{\bf Memory and storage defenses}
Several prior defenses considered offchip memory and storage outside of their trust boundary and proposed confidentiality and integrity protection.
\textit{Direct} attacks are thwarted by encrypting data in memory~\cite{tdx,sev} and storage~\cite{dm-verity}, preventing attackers from reading out plaintext secret.
Other works~\cite{costan2016intel, mee} introduced data integrity primitives with authenticated encryption blocks to prevent \textit{bitflip} attacks.
However, data integrity protection incurs latency and memory bandwidth overheads. 
Efforts to reduce the memory traffic overhead includes Morphable counters~\cite{morphable} that optimize the memory traffic by storing the message authentication code (MAC) in the ECC storage, while MGX~\cite{hua2022mgx} eliminates the need for loading replay counters by generating them in the processor.
Other works~\cite{aegisBitflip,saxena2022aqua,qureshi2022hydra,woo2023scalable} propose Rowhammer and other bitflip mitigation in memory.
\textit{Timing} channels are prevented by restructuring the data structures to prevent bank conflicts or defenses in memory controller~\cite{wang2014timing,ferraiuolo2016lattice,xiong2022secndp,shafiee2015avoiding}.
Finally, \textit{power} side channels are defended with masking and hiding techniques~\cite{sapui2023power}, while \textit{injection} attacks can be prevented with use of data integrity blocks.
It is important to monitor health of SSDs since these are used as storage buckets for training data and knowledge databases.
RL-Watchdog~\cite{rlwdog} introduces a reinforcement learning technique to detect SSD failures, while NVMensure~\cite{nvmensure} computes checksum to prevent data poisoning attacks during ML training.

\noindent{\bf Interconnect defenses}
IO information (including address, data, and timing) can leak information \cite{reverseCNN}. ORAM \cite{devadas2016onion} is a general way of protecting address and data from leaking information by obfuscating the address and the data content.
However, its high performance overhead prevents practical adoption. Practical secure accelerators perform encrypted data transfer to prevent \textit{direct} attacks~\cite{hua2022mgx,hua2022guardnn}. 
Bus hijack attacks can be prevented by securing DMA transfers with IOMMU~\cite{iommu} and RDMA protocol with sRDMA~\cite{srdma}.
Performance counters should be disabled with hardware fuses in production systems to prevent malicious hypervisors to snoop application data. 
Trusted execution environments like Intel TDX~\cite{tdx}, Nvidia CC~\cite{nvidia_cc}, or Arm Trustzone~\cite{trustzone} ensure that performance counters are disabled during trusted execution. 
The \textit{timing} channel with IO~\cite{invisiprobe} is not protected either by ORAM or encrypted data transfer.
There are currently no defenses to prevent sparsity data volume attacks. However, structured sparsity deployed in Nvidia Ampere GPUs~\cite{sparsity} can reduce the data leakage granularity.
The contention and fingerprinting attacks can be mitigated by memory traffic shaping~\cite{zhou2017camouflage,dagguise,banerjee2023triton}. 
This introduces high bandwidth utilization of ML accelerators, which is optimized by obsidian~\cite{obsidian}.
\textit{Bitflip} attacks with hardware trojans can be prevented by protecting the hardware supply chain~\cite{hwsupplychain}.
Introducing hardware trojans is easier in accelerator ASIC and FPGA implementations making it pertinent to verify FPGA bitstreams and attest accelerators to prevent bitflip due to malicious hardware.

\noindent{\bf Compute defenses}
The compute block defenses include on-chip memory, buffers and the processing unit. 
On-chip memory like cache and scratchpad leaks confidential information about the model or the input data via timing channel or power side channel. 
Cache timing side-channels is a well-studied problem with two types of defenses: 
Randomized cache designs like ScatterCache~\cite{scattercache} and others\cite{tan2020phantomcache,caesers}
insert secret data in random cache sets, while set partitioning approaches like DAWG~\cite{dawg} and others~\cite{townley2022composable,vantage} isolate cache ways. Both of these techniques can defend against neural network-specific attacks like \cite{cacheTelepathy}. 
Sesame~\cite{banerjee2020sesame} proposed scratchpad and on-chip buffer partitioning for ML tenants to protect against timing attacks.

Apart from memory structures, logical units and gates are also vulnerable to various physical and digital side-channels since compute logic are often optimized for specialized operand data. These optimizations are exploited to leak sensitive model weights and input data via timing measurements and power usages.
Power and EM variation can be measured by direct external physical equipments, while timing variations can be captured by both digital and physical probes. 
Power side-channels can be defended by specialized hardware unit that minimizes power variations \cite{reg_switch} or by obfuscating power signals using hardware masks similar to Bomanet~\cite{dubey2020bomanet}. Luo~\cite{powerdefense} proposed power side-channel mitigation combating voltage virus droop~\cite{majumdar2024voltage} in multi-tenant accelerators by reducing clock frequency with DVFS. Signal jamming techniques~\cite{jamming} are used to evade EM based fingerprinting~\cite{yu2020deepem} attacks. 

Several digital side-channels use performance counters and high-resolution timers to exploit optimized data-paths that leak model architecture or secret data values. Since 
trusted execution environments~\cite{tdx,nvidia_cc} are also susceptible to such vulnerabilities, they disable performance counters during confidential computation to mitigate side-channel analysis. 
Timing attacks can also be combated by eliminating high-resolution timers in production systems~\cite{rdtsc_resolution} or obfuscating timing measurement~\cite{martin2012timewarp}. 
Constant-timing hardware \cite{gleissenthall2019iodine,sakiyama2006small,aegis} provide microarchitectural mitigation by enforcing timing invariability in ALUs and removing operand data specific optimizations. Intel DOIT~\cite{doit} and ARM DIT~\cite{dit} in recent processors provide in specific instructions that enforce constant time executions and do not involve any data driven micro-architectural optimizations.

\subsection{Key Takeaways}
\textbf{\textit{Takeaway 1: Protecting the hardware supply chain:}}
Safeguarding the hardware supply chain is essential to prevent insertion of Hardware Trojans to leak secrets. 
Since, a lot of AI applications leverage multi-tenant FPGA accelerators, authenticating the bitstream is critical in preventing this powerful attack. 
Security verification of hardware micro-architecture~\cite{teesec} is critical to prevent side-channel attacks.

\textbf{\textit{Takeaway 2: Avoiding data-driven optimizations:}}
Timing variations in data-driven optimizations leak secret data. 
Data pruning in memory bus~\cite{reverseCNN}, sparse computation~\cite{sparsityTime} and cache data compression~\cite{compression} can leak secret data.
Such features should be disabled during secure computation.

\textbf{\textit{Takeaway 3: Optimizing hardware security primitives:}}
The challenge of high performance overheads in adopting obfuscation techniques can be addressed by adoption of 
domain-specific~\cite{hua2022mgx} and state-space exploration~\cite{secureloop,obsidian}. 
However, these optimizations should not result in new side-channels.
Emerging architectures like PiM and PnM can help reduce memory traffic and data movement latency.

\textbf{\textit{Takeaway 4: Creation of a hardware design template}}
Computation, data movement and storage are the important primitives in a compound AI system. 
TEEs with side-channel protections~\cite{banerjee2020sesame} can perform isolated execution, link encryption in I/O bus protects data confidentiality and reliable storage prevents data tampering.
Each hardware deployment should be follow a design template having a combination TEE compute unit, encrypted switches and reliable storage blocks, similar to the recent Nvidia confidential compute deployment~\cite{tdx_cc}.

\section{Cross-Layer Attacks}
\label{sec:systematize}

~\Cref{sec:sw_att_def} and~\ref{sec:hw_att_def} explore software and hardware vulnerabilities, along with available defenses for various threat models within a compound AI system. Here, we focus on cross-layer attacks, where adversaries exploit both software and hardware vulnerabilities to breach data security. In~\cref{subsec:collab}, we discuss existing cross-layer attacks that combine algorithmic, software, and hardware methods. We also highlight the potential severity of these attacks in modern compound AI systems, which are increasingly reliant on heterogeneous hardware and diverse software stacks.

As AI ecosystems evolve, the scope of these attacks must be reconsidered, especially since many traditional CPU-side attacks are being adapted to violate confidentiality, integrity, and availability in the AI/ML domain. Effective security cannot be achieved in isolation -- reliable protection requires a holistic approach that encompasses both ML algorithms and secure platform design, considering the entire system for threat modeling and defenses.

\subsection{Cross-layer attacks in AI model deployments}
\label{subsec:collab}
Systematizing a framework for existing proposed attacks, across the the application stack allows us to envision how \textit{algorithmic} attacks can be combined with \textit{software} and \textit{hardware} vulnerabilities to achieve higher efficacy. 
Many system attacks, especially those targeting hardware physical and digital channels, only protect part of the assets like model architecture or model weights. On the other hand, algorithmic attacks often make assumptions (e.g. whitebox or greybox) that are only feasible after partial model information extraction from system attacks.

Here we are presenting four cases that leverage both system and algorithmic techniques to mount powerful attacks. 
Each attack is mapped to MITRE techniques, as illustrated in~\cref{fig:collab_attack}, allowing for a clear classification of attack lifecycles. Such a mapping aids system designers in understanding the stages of an attack and selecting appropriate defense mechanisms to mitigate these threats. We discuss more on defense methodologies and quantifying attack techniques in~\cref{subsec:quants}.

\begin{figure}
    \includegraphics[trim={0.8cm 0.5cm 0.75cm 0.3cm},clip,width=1\linewidth]{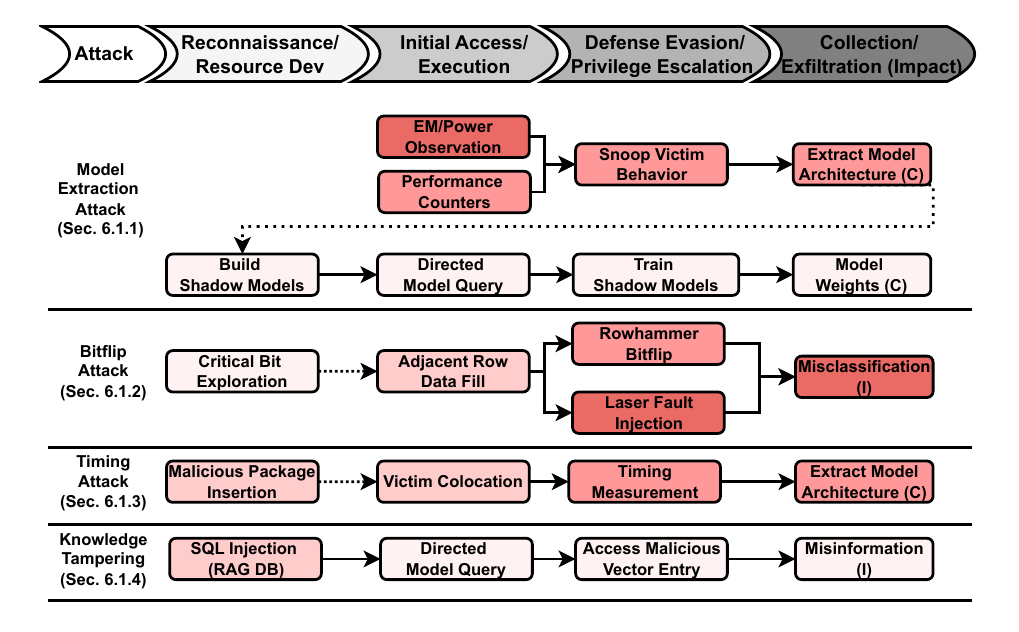}
    \caption{\textbf{Three attack cases which leverages cross-stack vulnerabilities to mount powerful attacks. We map different attack techniques to the MITRE framework.}}
    \label{fig:collab_attack}
\end{figure}

\noindent{\bf 6.1.1 Performing model extraction:}
An attacker with hardware physical access can extract model weights using coldboot attacks to directly read memory contents. 
However, attaining physical access to MLaaS deployments is infeasible.
The model extraction attacks in MLaaS threat models comprise to two steps: 
\textbf{(I)} Extract model layer architecture (filter matrix dimensions); 
\textbf{(II)} Train a duplicate model with same accuracy. 
Digital attacks like Deepsniffer~\cite{deepsniffer} and others~\cite{reverseCNN,band_util} monitor hardware utilization to reveal model architecture, while physical attacks like DeepEM~\cite{yu2020deepem} and others~\cite{opendnn} use EM and power channels.

Subsequently, an attacker can perform an algorithmic attack with whitebox assumptions.
The attacker creates a number of shadow models and train them on the repeated query response of the victim model. 
Ultimately, one of the shadow models will achieve the same level of accuracy as the victim model, resulting in a successful model extraction attack.
This attack requires only remote query access to the victim model.
This approach can also be used to perform membership inference attacks with a different asset target. 
While model extraction attacks target model parameter confidentiality ($M_v$), membership inference declassifies training data ($T$). 
In this example, we see how hardware attacks can transform weaker threat model assumptions of algorithmic attacks into a more realistic threat model.

\noindent{\bf 6.1.2 Bitflip for model degradation:}
While bit-flip attacks can be executed through various system attack methods, their impact varies significantly depending on the targeted bits. The effectiveness of such attacks can be amplified by following these steps:
\textbf{(I)} Identify the critical model bits that misclassify a specific class or degrades model accuracy. 
Proflip~\cite{proflip} and TBT~\cite{rakin2020tbt}, for instance explores the sensitive bits in a ML model.
\textbf{(II)} Mount a Deephammer~\cite{deephammer} attack to flip the specific bits for maximum impact. 
An attacker with physical access can also perform Laser fault injection~\cite{laserFltInj} to flip critical model bits. 
This approach highlights how algorithmic attacks can serve as reconnaissance tools, identifying critical bits to enhance the effectiveness of conventional bit-flip attacks.

\noindent{\bf 6.1.3 Covert channel for hardware timing attack:}
Hardware timing attacks like layer fingerprinting~\cite{fpmt, csi_nn} and cache side-channels~\cite{cacheTelepathy,deeprecon} are particularly effective when the attacker knows the start and end points of each layer's execution.
This helps eliminate noisy data collection from these side-channels.
An attacker can extract model architecture with only remote digital access, making this attack relevant in cloud MLaaS deployments.
This attack can be performed in two stages: 
\textbf{(I)} A malicious software package can be inserted in the software supply chain to trigger an interrupt at the compute start of each layer. 
\textbf{(II)} This covert channel can be intercepted by an attacker to reset or collect timing information for layer fingerprinting. 
The supply-chain attack denoises the timing attack data collection by the digital attacker for more precise fingerprinting. 
Moreover, a compromised orchestration framework can aid victim colocation exclusively with the attacker to eliminate system noise from other processes.

\noindent{\bf 6.1.4 Tamper knowledge database to spread misinformation:}
RAG models can generate recent content by retrieving from an updated knowledge database. 
ConfusedPilot~\cite{confusedpilot} shows how an attacker can spread misinformation by tampering the knowledge database. 
The key assumption is that an attacker can add malicious entries in the knowledge database. 
However, this might not be true for enterprise databases with administrator gate-keeping. 
Software attack can evade this assumption with the following attack flow:
\textbf{(I)} An SQL injection attack~\cite{cveF1.1} can insert malicious knowledge in the vector database. 
Such an attack was shown on Langchain framework (listed in~\cref{tbl:sw_att_list}) which is widely used in designing RAG pipeline and can be launched by an unprivileged attacker. 
\textbf{(II)} After tampering the database, the ConfusedPilot algorithmic attack can be deployed to tamper content generation integrity or RAG availability. 
This attack exemplifies how a specific threat model assumption in an algorithmic attack can be generalized with cross-layer attacks. 
Mapping these attacks to the Mitre framework help replace threat model assumptions in the reconnaissance and resource development stages with system attacks.

\subsection{Emerging cross-layer attacks on compound AI systems }

In this section, we will qualitatively discuss new cross-layer attacks possible on the compound AI system. 
Specifically, we showcase how an adversary can extract partial information from different compound AI stages (shown in~\cref{fig:AI-Pipeline} to design an end-to-end attack. 
Carlini~\cite{carlini} already showcased a purely algorithmic attack where a LLM model trained with differential privacy (DP) leaks training data violating privacy when deployed as a component in a compound AI system.
This seminal work shows how a safe training DP budget is violated, when a LLM is used as a part of a larger system.

\noindent{\bf Generation accuracy degradation:}
Data poisoning and other approaches are shown to degrade model accuracy or cause misclassification.
However, a compound AI system includes a grounding block (shown in~\cref{fig:AI-Pipeline}) in query post-processing to fact-check the generated content. 
This prevents model hallucination and the effectiveness of isolated data poisoning. 
An attacker with physical access can still generate a degraded response and bypass the grounding block to perform an attack.

The \textit{reconnaissance} step starts with finding the physical address of the multiplexer logic in the MoE router.
This can be performed by dumping device memory contents with memory safety vulnerabilities (use-after-free or OOB read) or by a hardware boot attack~\cite{coldboot}.
The \textit{initial access} stage involves installing a malicious grounding block to bypass the fact checking logic. 
Next, the attacker mounts a rowhammer attack to alter the MoE router expert selection for a victim query. 
This generates a degraded response which evades fact-check with a malicious grounding component completing the attack.
This attack requires hardware physical access, but is able to use cross-layer attacks to emit a degraded response despite query post-processing. 

\noindent{\bf Leak private data from generated program snippet:}
Compound AI tools like Github Copilot~\cite{gitCopilot} are widely used to generate program snippets.
Tampering the generated program is lucrative as it can be used to extract user private data used as program inputs, manipulate the desired output or crash a user system.

This attack starts by redirecting the LLM agent into a malicious program generator. 
This can be performed by modifying the function call address with a OOB write in the LLM agent. 
The malicious program generator can add covert channel attack widgets in the generated code. 
When the victim executes the generated code snippet, it triggers the covert channel enabling the attacker to leak secret data with cache~\cite{flushReload} and other hardware timing~\cite{nakai2021timing} attacks.

\noindent{\bf Leak model parameters with Hardware Trojan:}
Compound AI system is composed of multiple proprietary LLMs. 
We showcase the model extraction attack by an attacker with physical access.
The \textit{reconnaissance} step includes insertion of an FPGA accelerator in the orchestration pool with a Hardware Trojan to extract model weights.
The privileged attacker hijacks the software orchestrator to redirect the execution of the target model into the accelerator.

Once the target model is scheduled in the malicious accelerator, the trojan extracts model parameters by snooping on the memory interconnect hardware channel. 
This attack can be generalized to leak any component in the compound AI system, including the retrieval output, LLM agent model, the MoE experts or the compliance and grounding LLMs.

\noindent{\bf Privilege escalation attack to control hardware:}
The privilege escalation attack is showcased as a software attack in~\cref{subsec:sw_att}.
However, the diverse software supply chain in a compound AI stack makes it more compelling. 
The ever-increasing use of packages and libraries make it more prone to bugs and malicious package inclusions. 
This attack is powerful as the attack only needs query access to perform privilege escalation. 
SSRF attacks in Langchain~\cite{cveF5.1} or access control violations in Langflow~\cite{cveF5.2} and Ollama~\cite{cveF5.3} increases the attacker capability. 
Mounting privilege escalation attack in the \textit{initial access} sets the stage for 
better control of the deployment platform. 
It enables a non-privileged attacker to toggle performance counters leading to hardware attacks or control the victim execution to colocate with the attacker or trigger DoS attacks.

\section{Discussion}
\label{sec:discussion}

In this SoK, we classify attacks on AI systems into three categories: algorithmic, software, and hardware attacks. 
We also demonstrate how cross-layer vulnerabilities can either amplify certain attacks or enable a less capable adversary to carry them out.
In this section, we first highlight the key takeaways from this exercise in~\cref{subsec:learnings}.
These learning help in designing a more holistic compound AI platform, that is robust at every level of stack. 
Next, stress the need of qualitatively and quantitatively analyze attack vectors in~\cref{subsec:quants}.
Since, systems have a diverse threat model assumption, understanding the severity of individual vulnerability is necessary to understand the severity of attacks and work towards quantifiable defenses.
Finally, we discuss various open research challenges in designing robust AI systems in~\cref{subsec:dis_defense}.

\subsection{Key Takeaways}
\label{subsec:learnings}
\textbf{\textit{Takeaway 1: The potency of cross-layer attacks:}}
The vulnerabilities in \textit{algorithmic}, \textit{software} and \textit{hardware} equip an attacker with different attack widgets.
These widgets can be sequenced to launch an end-to-end attack on a compound AI system with minimal threat model assumptions.
Multiple ownership of the hardware platforms and the software supply chain makes it challenging to propose a single defense. 
Software and hardware defenses are needed at every stage of the training and inference of compound AI pipelines.

\textbf{\textit{Takeaway 2: The need for a holistic attack categorization:}}
Given the diverse asset and threat model categories for isolated attacks, there is a need for a holistic attack categorization. 
We propose using the $Mitre\ Att\&ck$ framework to position different attacks based on the threat model, the leaked asset and the overall impact. 
While isolated attacks like few bitflips may be harmless in attacking a billion parameter model, it can be lethal if paired with an algorithmic attack to flip a few critical bits. 
A severity score should be assigned to isolated attacks based on the threat model, the type of leaked asset and the overall impact on a large system like in compound AI.

\textbf{\textit{Takeaway 3: Identifying the critical attack targets:}}
The \textit{reconnaissance} step enables the attacker to identify critical stack component that should be targeted for a specific secret asset. 
For instance, knowledge databases and grounding blocks are the critical components, if the attacker wants to degrade response accuracy or spread misinformation. 
An attacker with physical access should flip bits at the later stages, as compound AI system is robust enough to correct early stage bitflip in many cases.
The LLM agent plays a crucial role in determining the compound AI’s response pathway, making it a high-value target for control-flow hijacking. 
Targeted model extraction and membership inference attacks needs to be directed at specific IPs. A malicious orchestrator can enable an attacker to colocate with the target component or deploy it in a specific node. 
These insights enable the attacker to collect noise free data during exfiltration stages to perform efficient attacks.



\textbf{\textit{Takeaway 4: Re-evaluation of existing defense mechanisms:}}
Many of the defense mechanisms are focused on a specific security policy and are deployed for a specific hardware platform. 
For instance, buffer overflow and memory safety protections are implemented in CPUs. 
However, compound AI systems are deployed in a heterogeneous hardware. 
Such protections should either be triaged to GPUs and accelerators, or devices should come with a defense capability metadata verified during attestation. 
This can help the software orchestrator to schedule sensitive components only to the secure systems. 
Moreover, supply chain defenses are becoming super important and challenging, given the dynamism of capability development in the compound AI systems.
Security policies of every software and hardware components should be designed based on first principle. 

\subsection{Need for Quantitative and Qualitative Metric}
\label{subsec:quants}

The cybersecurity and software industries have utilized quantification metrics, such as CVEs, to assess the severity of attacks and vulnerabilities. While hardware vulnerabilities lack this quantification, metrics like channel capacity and leakage bit-rate have been proposed in the literature~\cite{casen,casa} to evaluate side and covert channels. Weaknesses in both software and hardware are categorized under Common Weakness Enumeration (CWE) metrics, enabling the community to identify and mitigate security violations before exploitation.
However, a classification mechanism for AI/ML attacks across the stack is currently lacking. Critical vulnerabilities in the components of compound AI systems can jeopardize the entire system's security guarantees. Although some groups are advocating for this issue~\cite{mitre-aiwg}, we lack significant outcomes. Rapid emergence of AI inference pipelines highlights the necessity of developing standardized metrics to categorize and quantify attacks at each layer.

We also call for a severity scoring system based on the sensitivity of targeted assets and the capabilities required by potential attackers. For example, while both model parameters and architectures are proprietary, leaking model parameters poses a greater sensitivity risk than model architecture. Additionally, different types of hyperparameters vary in their criticality regarding model accuracy, which must be factored into severity assessments.
Beyond asset sensitivity, classifying threat models and attacker capabilities, plays a crucial role in determining attack potency. For instance, vulnerabilities exploitable by remote adversaries have a broader attack "blast radius" than those requiring physical access. Similarly, an algorithmic attack with white-box access is typically less potent than one requiring only black-box access.
This work, organizes existing attacks into a recognized framework that aids in classifying AI vulnerabilities. Extending the framework with quantifiable measures is an open research question we aim to address in future.

\subsection{Building a Robust Compound AI System}
\label{subsec:dis_defense}

Here we explore best practices for designing secure compound AI systems that aim at cross layer vulnerabilities and employ defenses at the algorithm, software, and hardware.

\noindent{\bf Cross-stack alignment of security requirements:}
Modern AI pipelines handle sensitive data from multiple distrustful entities, complicating security enforcement. Privacy policies from end-users and security requirements from model owners often lack clear alignment with software and hardware layers, making it challenging to maintain consistent security guarantees.
Establishing clear trust relationships between data owners and entities, along with specifying privacy requirements in multi-tenant environments, is crucial for designing robust AI systems.
We identify the definition of trust relationships and the creation of a security framework that can effectively translate user policies into system guarantees as key open research questions for future work.

\noindent{\bf Factoring DP budget in a compound AI system:}
Differential privacy (DP)~\cite{dwork2006differential} is a mathematical tool to quantify the privacy budget. 
DP stochastic gradient descent~\cite{dpsgd} is already used to train LLMs.
However, the DP privacy budget should account for other components in the compound AI pipeline.
The compositional nature of DP that the total privacy budget equals the sum of privacy budget of each component makes it convenient to quantify the privacy for complex systems like compound AI system. 
DP-based framework for quantifying privacy leakage and providing guidance towards protecting layer-designs in a compound AI systems is an open area for future studies.

\noindent{\bf Cross-layer IFC for compound AI:}
Information flow control (IFC) can help prevent data leakage in compound AI systems by tagging confidential data in vector databases, restricting sensitive information from reaching malicious entities. Prior work has applied IFC to control expert access in MoE models \cite{tiwari2024information}, and this approach can be extended to compound AI systems. Enforcing IFC early in the pipeline limits data leakage, while memory safety measures protect against access control evasion and code injection. Metadata like ownership and permissions on database entries can enable fine-grain access control for sensitive data. Current IFC methods focus on algorithms but overlook system-level attacks. Cross-layer IFC, including hardware and software\cite{ferraiuolo2018hyperflow, tiwari2011crafting, banerjee2023triton} and full-system verification \cite{erbsen2021integration}, offers stronger security and is a promising direction for AI security.

\noindent{\bf Private Attested AI Services:}
Recent AI service offerings Apple's Private Cloud Compute (PCC)~\cite{pcc} and Edgeless System's Continuum AI~\cite{edgelesscontinuum}
start a trend to place AI services into a highly confined environment. In addition, these systems offer users attestation evidence of 
the environment which increases trust into these systems. Confidential computing (CC) like Intel TDX~\cite{tdx} provides the foundation to
provide this evidence to remote parties and in combination with trusted supply chains~\cite{sw_supply_chain} can further strengthen
the trust in the entire deployment. CC furthermore reduces the reliance on a trustworthy infrastructure provider, when the trusted execution environment
is implemented by a 3rd party (unlike Apple's PCC).

\noindent{\bf Identifying critical model parameters:}
Recent studies\cite{rakin2020tbt,aegisBitflip,proflip} have demonstrated that some ML model parameters are more critical than others. 
For instance, multiplexer bits in a MoE router or bits in a RAG indexing algorithm can significantly increase misclassification risks in a compound AI system. 
Similar analyses are needed for other components, including LLM agents, to identify key parameters. 
Quantifying critical parameters is essential for designing defense mechanisms and differential privacy budgets. 
Such a study should also include a sensitivity analysis of random bit flips to assess their impact on model accuracy, helping specify toleration threshold for deployments

\section{Conclusion}
\label{sec:conc}
This SoK systemizes the software and hardware vulnerabilities to forecast cross-layer vulnerabilities in a compound AI stack. 
While existing attacks and defenses have diverse threat models, we strive to systemize them using a well-known $Mitre\ Att\&ck$ framework. 
Arranging attack widgets according to the assumed threat model and target assets help clarify the flow of an attack sequence.
It further sets the stage for quantitatively scoring the potency of different vulnerabilities.
This paper highlights the risk of system attacks in the software and hardware of machine learning systems. 
This SoK lays the foundation for future advancements in both attack and defense strategies, expanding focus beyond algorithmic attacks to include vulnerabilities across software and hardware stack layers.

\section{Acknowledgment}
We thank Carlos Rozas, Mona Vij, Cory Cornelius, Scott Constable, Mic Bowman, Nageen Himayat, Jose Vicarte, Marius Arvinte, Fangfei Liu, Sebastian Szyller and other Intel SPR team members for the regular discussions and valuable feedback. 
This work was supported in part by ACE, one of the seven centers in JUMP 2.0, a Semiconductor Research Corporation (SRC) program sponsored by DARPA.

\bibliographystyle{ieeetr}
\bibliography{ML_SoK_arXiv/SoK-arXiv}


\end{document}